\documentclass[10pt]{ismd08}
\usepackage{graphicx,epsfig,floatflt}
\usepackage{cite,./mcite}

\newcommand{\pom}{I\!\!P}

\begin{document}
\title{Inclusive diffraction and factorisation at HERA}
\author{M.~Wing$^1$
}
\institute{$^1$University College London, DESY and Universit\"{a}t Hamburg}
\maketitle
\begin{abstract}
In this article, recent measurements of diffraction in deep inelastic scattering 
are presented along with QCD fits to extract the partonic structure of the 
exchange.  These so-called diffractive parton density functions can then be used 
in predictions for other processes to test factorisation in diffraction.  This 
is an important verification of QCD and has significance for predicting 
exotic signals such as diffractive Higgs production at the LHC.
\end{abstract}

\section{Introduction}
\label{sec:intro}

Diffraction in deep inelastic scattering (DIS) has long been a subject of 
great interest since the discovery of the first striking events at the 
beginning of the HERA programme~\cite{pl:b315:481,np:b429:477}.  The final state of a 
diffractive $ep$ collision at HERA contains a high energy scattered electron 
measured in the detector and a proton which remains intact and exits through 
the beam-pipe, sometimes to be detected in proton spectrometers along the 
proton beam-line.  
In addition the event consists of hadronic activity in the main detector, but 
with none in the direction of the proton.  This dearth of hadronic activity 
in the proton direction constitutes the striking experimental signature 
which caused great surprise in the early years.  Along with this so-called 
large rapidity gap (LRG), the hadronic final state has a very low invariant 
mass, $M_X$, compared to non-diffractive DIS.  All three signatures are used 
to isolate diffractive events. The techniques complement each other with detection of 
the final-state proton providing the cleanest signature but also with much 
lower statistics and a more restricted kinematic range.  The LRG and $M_X$ 
methods are similar in range and statistics but have different background 
contaminations.

\begin{floatingfigure}[r]{5cm}
\centerline{\epsfig{file=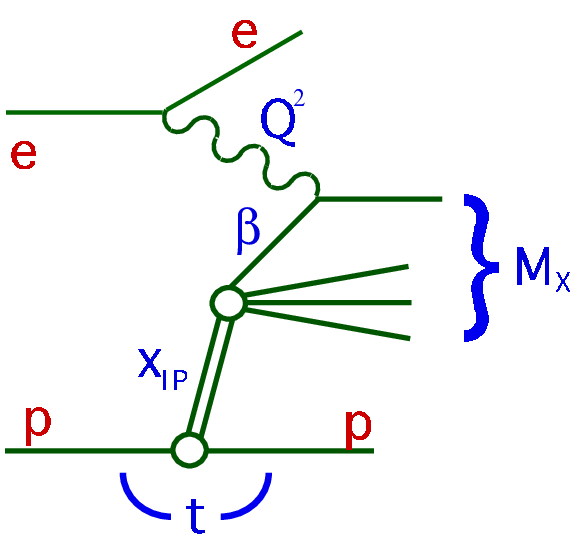,height=4.5cm}}
\caption{Schematic of diffraction in DIS.}
\label{fig:feynman}
\end{floatingfigure}
Such events can be understood in terms of the exchange of a colourless object, 
sometimes known as the Pomeron, which develops a structure.  The virtual photon 
emitted from the electron collides with a parton in this colourless 
object producing a hard collision. 
Figure~\ref{fig:feynman} shows this process along with relevant kinematic 
variables.  The cross section for diffractive processes can 
be factorised into the convolution of the Pomeron flux, $f_{\pom}$, as suggested by 
Regge theory, diffractive parton density functions (dPDFs), $f_{i/\pom}$,  and 
the hard scatter between one of the partons from the diffractive exchange and the photon, 
$\sigma_{ep \to eXp} \sim f_{\pom} \otimes f_{i/\pom} \otimes \sigma_{i \gamma \to jk}$.

There are several motivations to study the nature of diffractive processes and 
learn more about QCD, {\em viz}: diffractive processes constitute a large 
fraction of inclusive cross section; the transition from ``soft'' to ``hard'' 
regimes~\cite{wegener}; the applicability of the factorisation approach; and 
the potential for major discoveries such as the Higgs boson produced in 
diffractive processes at the LHC which relies on the above understanding.

This article reviews the most recent measurements of inclusive diffraction in 
DIS and the extraction of dPDFs from such data.  Factorisation is then tested 
through comparison of dPDFs (convoluted with an appropriate programme to calculate 
the hard scatter) for jet production in DIS and photoproduction as well as 
at the Tevatron.

\vspace{-0.3cm}
\section{Inclusive diffraction in DIS}
\label{sec:inclusive}

The ZEUS collaboration has recently published results on inclusive diffraction 
in DIS using all three methods~\cite{np:b800:1,zeus-pub-08-010}.  The data from 
the $M_X$ method extend the previous results~\cite{np:b713:3} to higher photon 
virtuality, $Q^2$, and, in the region of overlap, with increased precision.  The 
data using the LRG method is a significant update over previous ZEUS measurements 
with this method.  It covers the same kinematic range as the data from the $M_X$ 
method and complements the previously released data using the LRG method from the 
H1 collaboration~\cite{epj:c48:715}.  Similarly, the data where the proton is 
tagged using the leading proton spectrometer (LPS) complement previous 
measurements~\cite{epj:c38:43,epj:c48:749}.

\begin{figure}[htp]
\begin{center}
\includegraphics[height=7.5cm]{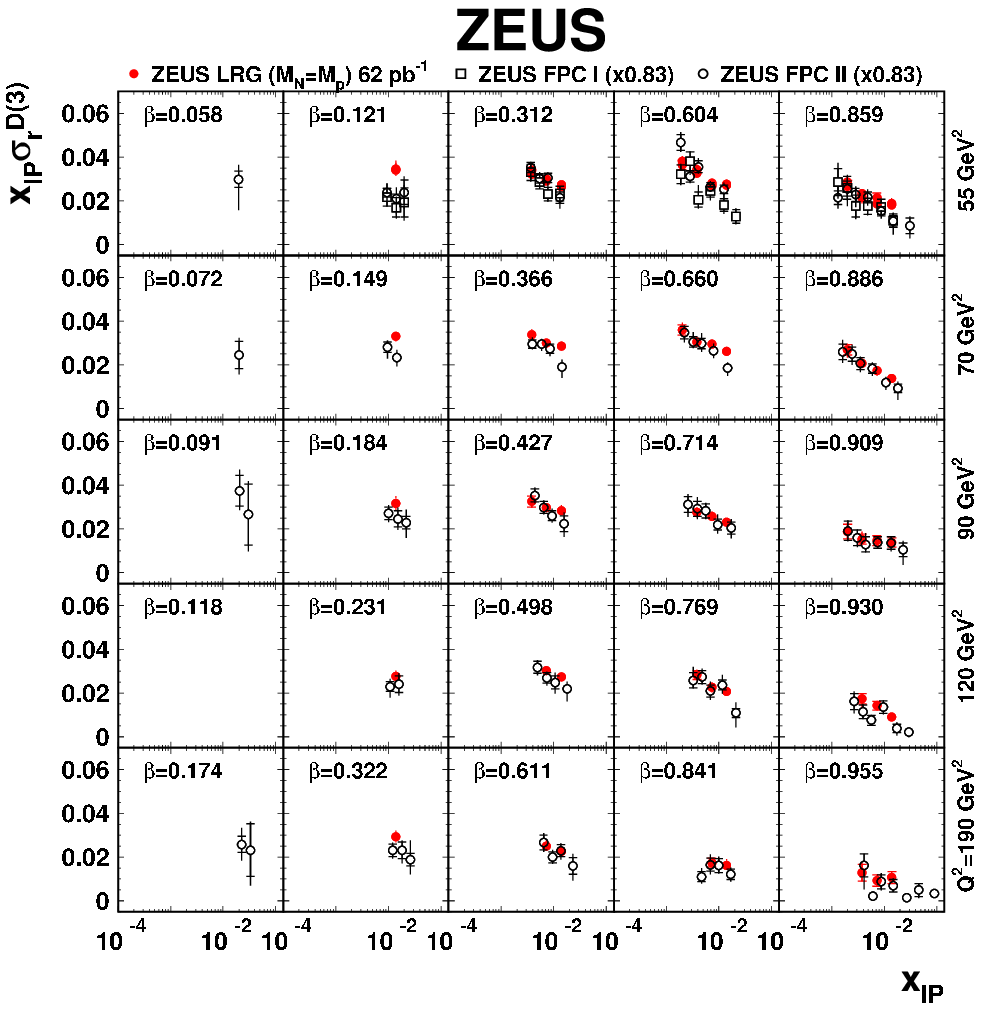}
\includegraphics[height=7.5cm]{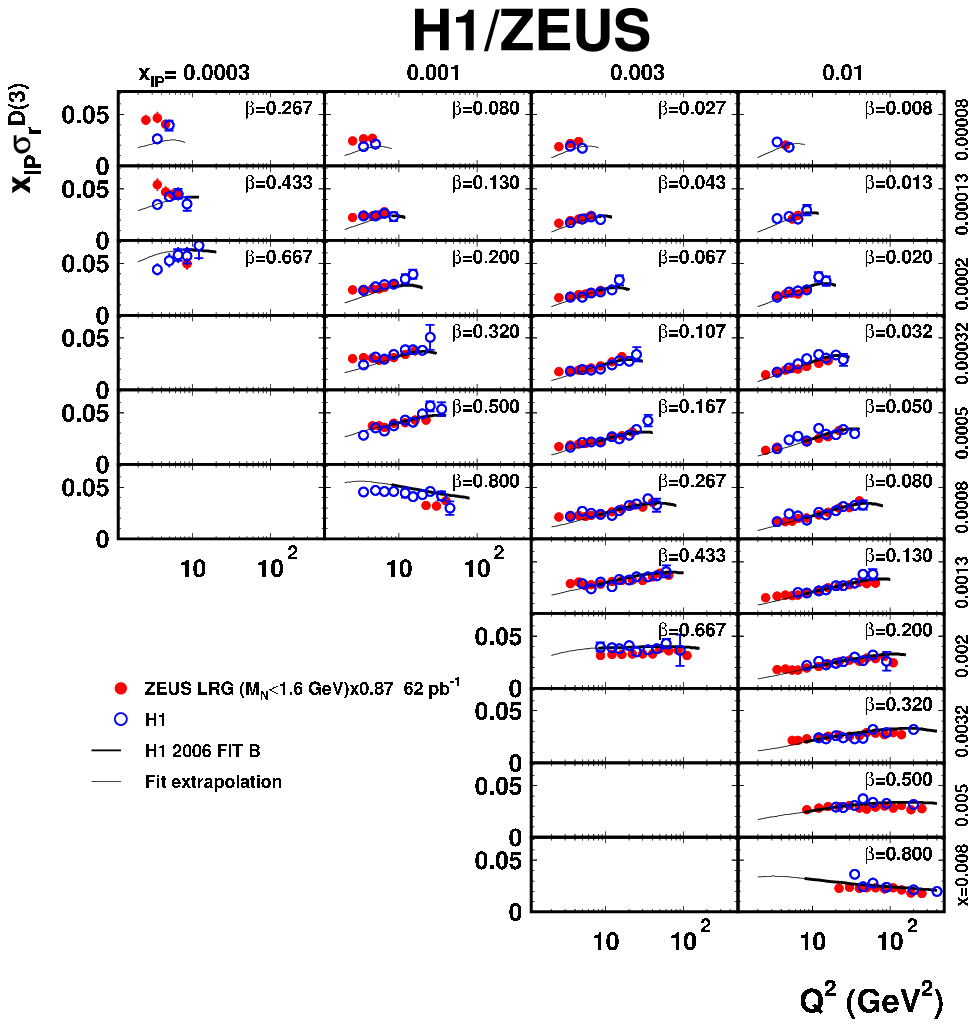}
\end{center}
\vspace{-0.5cm}
\caption{Comparison of the reduced cross section in inclusive diffractive DIS as a 
function of (left) $x_{\pom}$ for fixed $\beta$ and $Q^2$ for the LRG and $M_X$ 
methods and (right) $Q^2$ for fixed $x_{\pom}$ and $\beta$ for H1 and ZEUS 
data using the LRG method.}
\label{fig:inclusive}
\end{figure}

A comparison of the ZEUS measurements of the inclusive reduced cross section 
using the LRG and $M_X$ methods is shown in Fig.~\ref{fig:inclusive}~(left) for 
the high-$Q^2$ data as a function 
of the Pomeron momentum fraction, $x_{\pom}$, at fixed $Q^2$ and fixed Pomeron 
momentum fraction carried by the parton in the hard scatter, $\beta$.  The data 
using the $M_X$ method are scaled to account for the residual 
background from proton dissociation in which a proton breaks up into a low-mass 
nucleon. Some 
differences between the methods (more marked at lower $Q^2$, not shown) as a function of $x_{\pom}$ 
are observed which can be attributed to the suppression of Reggeon and pion 
trajectories at high $x_{\pom}$ in the $M_X$ method.  Also 
at lower $Q^2$, the two measurements have a somewhat different $Q^2$ dependence 
with the data from the $M_X$ method decreasing faster with the $Q^2$ than those 
from the LRG data.  However the overall agreement between the two data sets is reasonable.

The measurements of the reduced cross section from both H1 and ZEUS collaborations 
using the LRG method are compared in Fig.~\ref{fig:inclusive}~(right) as a function 
of $Q^2$ for fixed $x_{\pom}$ and $\beta$.  To enable a comparison in shape, the 
ZEUS data have been normalised to the H1 data within 
the uncertainty in the relative normalisation of the two measurements.  Overall 
the (qualitative) agreement is good and work is ongoing to combine the measurements 
which will give a quantitative measure of their compatibility and possibly lead to 
a significantly improved determination of the cross section.  
Already from these data it can be seen that at fixed $\beta$, the $Q^2$ dependence 
is different for different $x_{\pom}$ values.  This effect, also seen in the results  
using the $M_X$ method, means that the data cannot be described by a single 
factorisable Regge contribution, $f_{\pom}$.  

Results in which a forward-going proton was tagged not only provide a clean measure 
of diffraction, but also as a result allow the determination of the residual 
background from proton dissociation, which is independent of all kinematic variables, 
in the other data samples.  This and other results of these measurements are 
discussed in the relevant publications~\cite{epj:c38:43,epj:c48:749,zeus-pub-08-010}. 

\vspace{-0.3cm}
\section{Extraction of dPDFs}
\label{sec:dPDFs}

\begin{floatingfigure}[r]{7cm}
\centerline{\epsfig{file=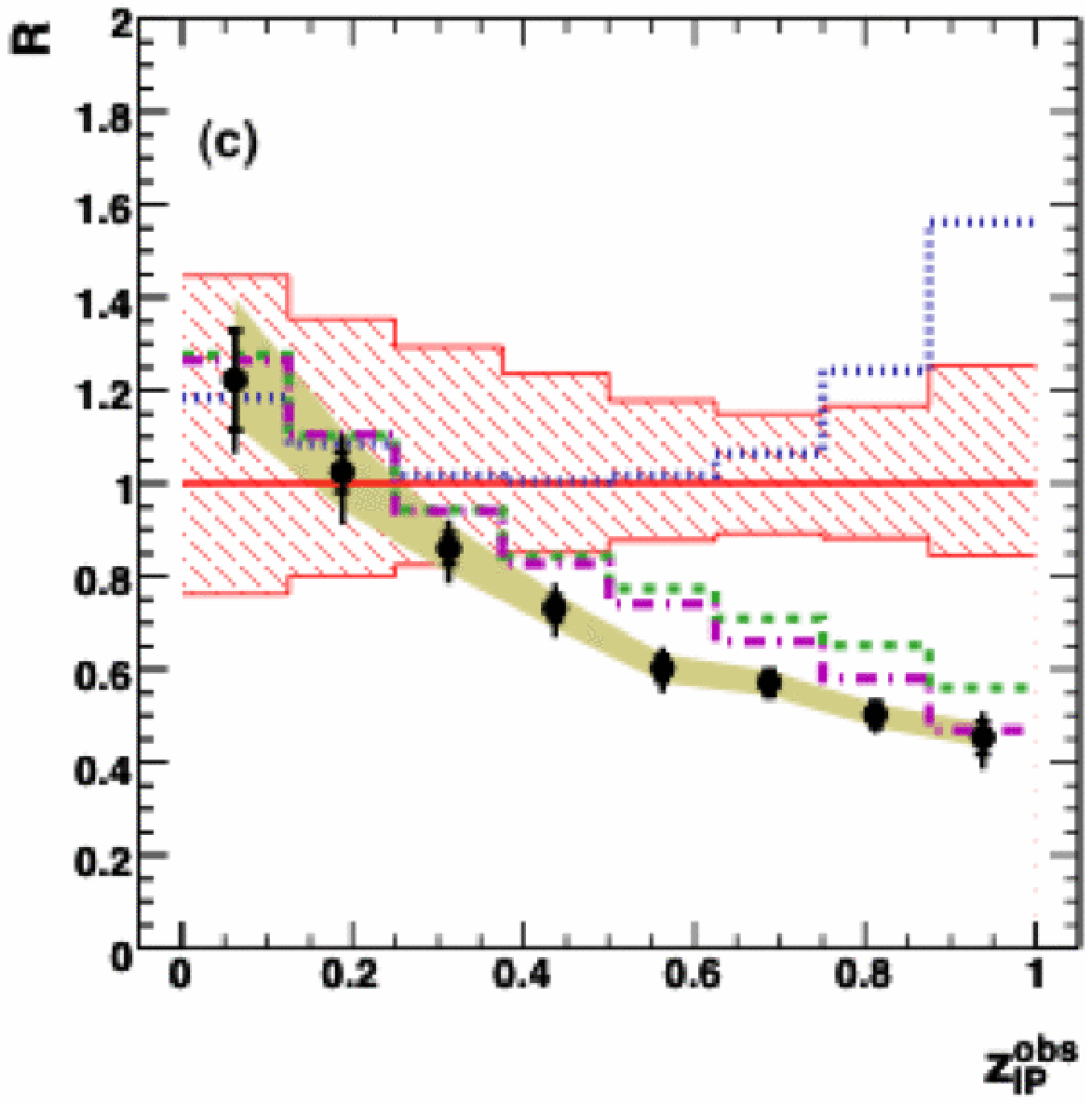,height=6cm}}
\vspace{-0.2cm}
\caption{Ratio of ZEUS data to NLO QCD theory as a function of $z_{\pom}^{\rm obs}$ 
for different dPDFs: a ZEUS fit to LPS and 
charm data (solid line); a H1 fit to inclusive LRG data, H1 fit 2006 - A (dotted 
line) and H1 fit 2006 - B (dashed line); and a fit to inclusive data from Martin 
et al., MRW 2006 (dot-dashed line).}
\label{fig:jets}
\end{floatingfigure}
The H1 collaboration pioneered fits in next-to-leading-order (NLO) QCD to the dPDFs.  
The inclusive data presented in the previous section was fit~\cite{epj:c48:715} 
and found to be 
dominated ($\sim$70\%) by the gluon density in the diffractive exchange.  However at large 
longitudinal momentum fraction, $z_{\pom}$, of the parton relative to the diffractive 
exchange, the data lack constraining power.  Although the quark contribution 
is stable, the gluon density can vary considerably when choosing different 
parametrisations.  This residual uncertainty (larger than other theoretical and 
experimental uncertainties) needed further input and was reduced by considering 
jet production in DIS and simultaneously fitting~\cite{jhep:0710:042} these and 
the inclusive data.

Figure~\ref{fig:jets} shows data from ZEUS on jet production, 
similar to that used by H1 in the NLO QCD fit for the dPDFs.  The ratio of the 
measured cross section to NLO QCD predictions with different dPDFs is 
shown as a function of the experimental estimator of $z_{\pom}$.  The data show 
clear sensitivity to the choice of dPDF with the theoretical predictions differing 
by up to a factor of 3 coming from the weak constraints on the gluon density.  There is also a clear preference for two of the dPDFs, 
MRW~2006 and H1 fit 2006 B, where the latter is one of the above two parametrisations 
derived from fits to inclusive data.  The good description of the data by these two 
parametrisations also demonstrates the applicability of factorisation in diffractive 
DIS.  These results demonstrate that jet data can be used in NLO QCD fits to 
further constrain the dPDFs.  

\begin{figure}[htp]
\begin{center}
\epsfig{file=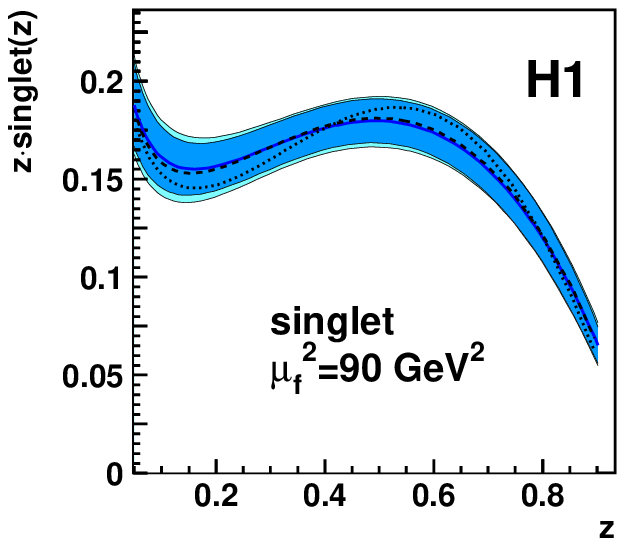,width=6.1cm}
\epsfig{file=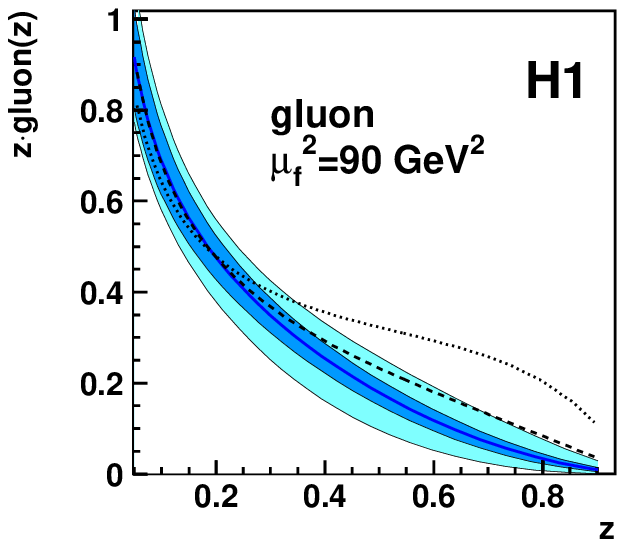,width=6.1cm}
\epsfig{file=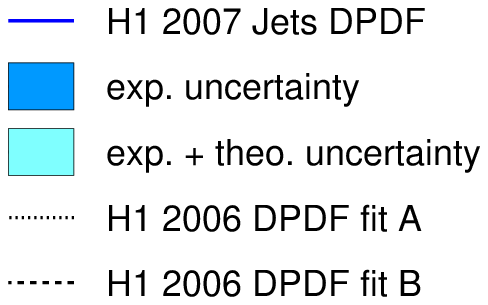,width=2cm}
\caption{Comparison of dPDFs for the quark and gluon densities when simultaneously 
fitting inclusive and jet data (bands) and when fitting inclusive only (lines).}
\label{fig:dPDFs}
\end{center}
\end{figure}

An NLO QCD fit was performed for the jet data as a function 
of the variable $z_{\pom}$ at different scales and in combination with the inclusive 
data.  The resulting parton densities for quark and gluons are shown in 
Fig.~\ref{fig:dPDFs} in comparison to the previous fits to the inclusive data 
only.  The new parametrisation of the gluon density follows that of H1 fit 2006 B 
and is now similarly well constrained in comparison with the quark density over 
the whole kinematic range.  

\vspace{-1cm}

\section{Diffractive jet photo/hadroproduction}
\label{sec:jets}

\begin{figure}[htp]
\begin{center}
\epsfig{file=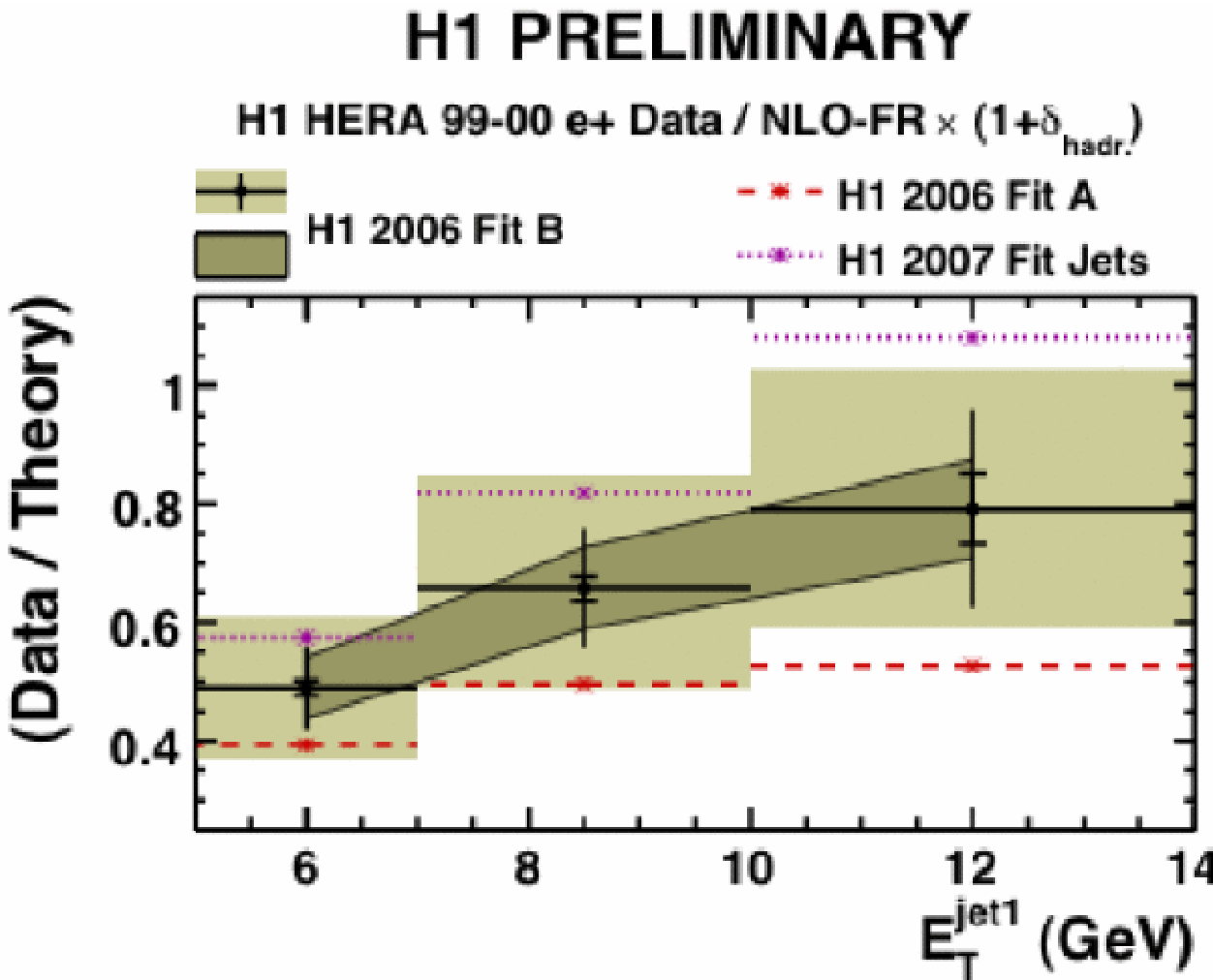,width=6.1cm}
\epsfig{file=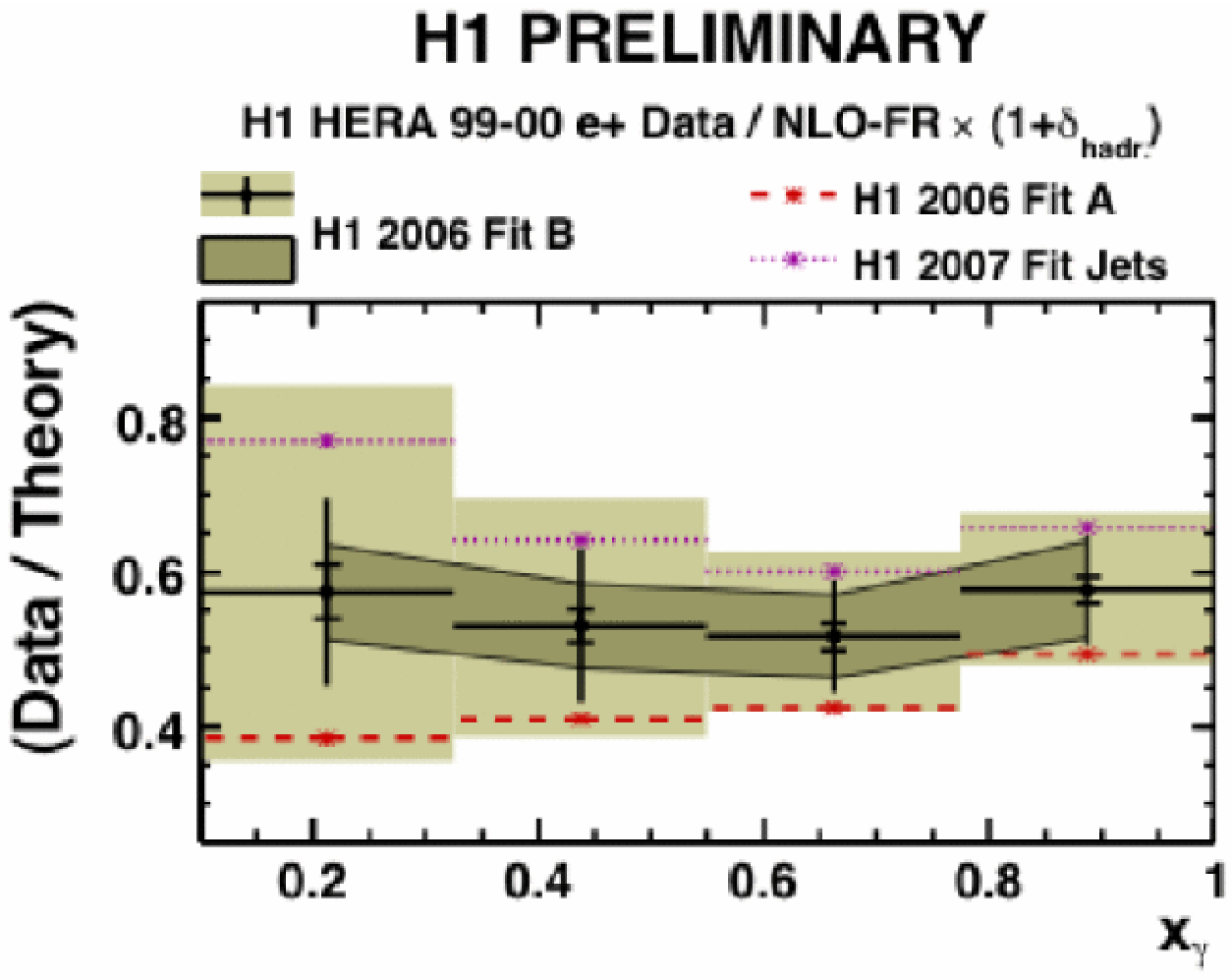,width=6.1cm}
\caption{Ratio of data to theory in jet photoproduction as a function of 
$E_T^{\rm jet1}$ and $x_\gamma$.}
\label{fig:gamma-p}
\end{center}
\end{figure}

It has long been observed that when dPDFs are compared to Tevatron 
data~\cite{prl:84:5043}, the rate 
is overestimated by about a factor of 10.  Explanations of this factorisation 
breaking exist~\cite{pl:b559:235} which predict secondary (multiple) interactions between 
the remnants which destroy the rapidity gap signature of diffraction.  It 
might also be expected for this to occur in photoproduction in which the 
almost-real photon develops a structure and can effect a hadronic collision.  
A useful variable to isolate such interactions is $x_\gamma$ which is the 
fraction of the photon's momentum participating in the hard scatter.  High 
values, the direct process, indicate the photon was point-like whereas lower 
values, the resolved process, indicate that the photon developed some structure.  
However, as can be seen in Fig.~\ref{fig:gamma-p} and also confirmed by ZEUS 
data~\cite{epj:c55:177}, no dependence of a suppression factor is seen as a 
function of $x_\gamma$.  There are indications of an overall suppression factor 
which (also) depends on the jet transverse energy.

\vspace{-0.3cm}
\section{Discussion}
\label{sec:discussion}

At first sight the situation in photoproduction and hadroproduction seems 
contradictory.  However, it should be noted that the nature and rate of 
secondary interactions in the two processes is almost certainly different.  
From inclusive jet photoproduction data~\cite{epj:c1:109}, secondary interactions 
are expected, but almost certainly not at the same rate as in hadroproduction.  
It should be remembered that in photoproduction, part of the resolved collisions 
look like the collision of a structured, vector-meson like, object with a proton.  
However, there is also the perturbative point-like splitting of the photon,  
which is fully calculable in QCD~\cite{np:b120:189}, in which the photon is not 
a structured object in the same way as for the vector meson model.  This is in 
contrast the obvious structured objects in hadroproduction.

\vspace{-0.5cm}
In summary, new measurements of inclusive diffraction have been made and new 
determinations of the partonic structure of the diffractive exchange calculated.  
These new parton densities demonstrate the applicability of factorisation in deep 
inelastic scattering, but do not change the situation in hadroproduction where 
models of secondary interactions are invoked to alleviate this breaking of 
factorisation.  The situation in photoproduction is less clear cut, but also 
does not contradict the results in DIS or in hadroproduction.  Further improvements 
will be made with the analysis of more inclusive and jet data and combination 
of data sets from the two collaborations.

\vspace{-0.3cm}
\section*{Acknowledgements}

M. Albrow, M. Diehl and J. Chyla are acknowledged for their questions which 
contributed to the above discussion.  The Alexander von Humboldt Stiftung is 
also gratefully acknowledged.

\vspace{-0.3cm}
\begin{footnotesize}
\bibliographystyle{ismd08} 
{\raggedright
\bibliography{ismd08}
}
\end{footnotesize}
\end{document}